# Vernier caliper and micrometer computer models using Easy Java Simulation and its pedagogical design feature-ideas to augment learning with real instruments

Loo Kang WEE[1], Hwee Tiang NING[2]

[1]Ministry of Education, Educational Technology Division, Singapore
[2]Ministry of Education, National Junior College, Singapore
wee_loo_kang@moe.gov.sg, ning_hwee_tiang@moe.edu.sg

Abstract: This article presents the customization of EJS models, used together with actual laboratory instruments, to create an active experiential learning of measurements. The laboratory instruments are the vernier caliper and the micrometer. Three computer model design ideas that complement real equipment are discussed in this article. They are 1) the simple view and associated learning to pen and paper question and the real world, 2) hints, answers, different options of scales and inclusion of zero error and 3) assessment for learning feedback. The initial positive feedback from Singaporean students and educators points to the possibility of these tools being successfully shared and implemented in learning communities, and validated. Educators are encouraged to change the source codes of these computer models to suit their own purposes, licensed creative commons attribution for the benefit of all humankind.

Video abstract: http://youtu.be/jHoA5M-_1R4
2015 Resources: http://iwant2study.org/ospsg/index.php/interactive-resources/physics/01-measurements/5-vernier-caliper
http://iwant2study.org/ospsg/index.php/interactive-resources/physics/01-measurements/6-micrometer

PACS: 06.30.Gv 06.30.Bp 1.50.H- 01.50.Lc 07.05.Tp

## I. INTRODUCTION

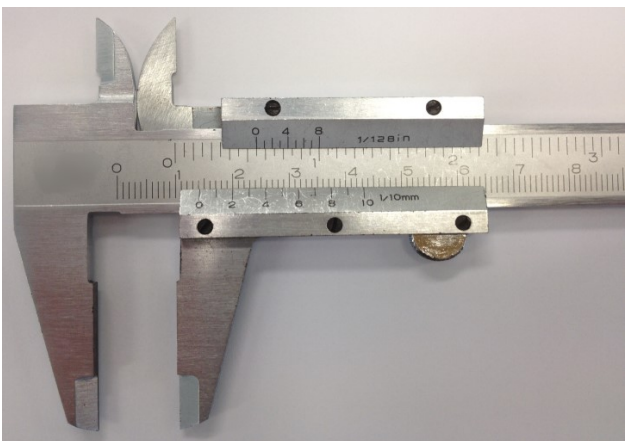

Figure 1. A vernier caliper used in the school and the vernier bottom scale spans 29 mm of the top scale. The reading is 14.2 mm.

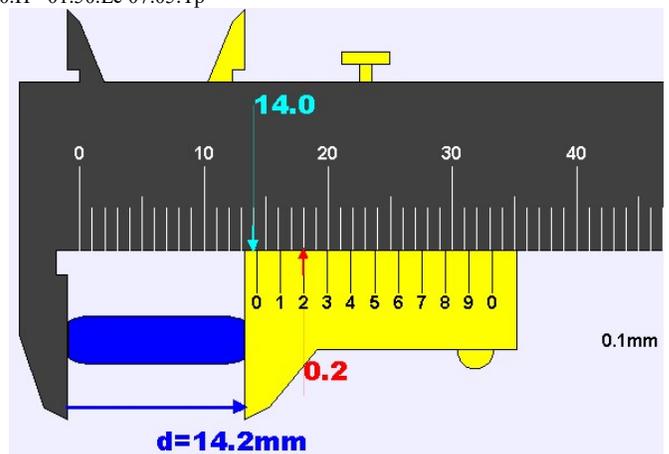

Figure 3. An EJS vernier caliper computer model with vernier bottom (yellow) scale that spans 29 mm of the top scale (darkgray). The hint shows 14.0 mm on the top scale, 0.2 mm on the bottom scale, giving the final answer of 14.2 mm. This model is downloadable
https://dl.dropbox.com/u/44365627/lookangEJSworkspace/export/ejs_AAPTVernierCaliper.jar

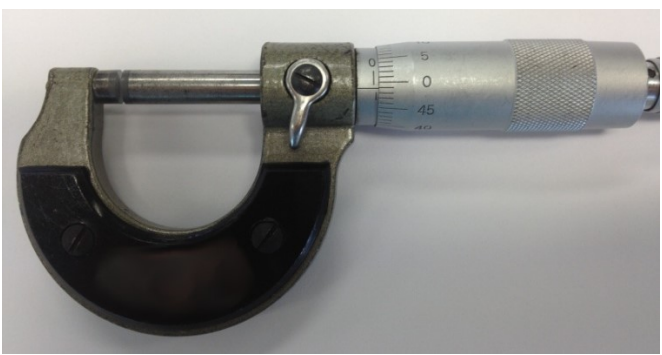

Figure 2. A micrometer used in the school and the reading is 0.99 mm.

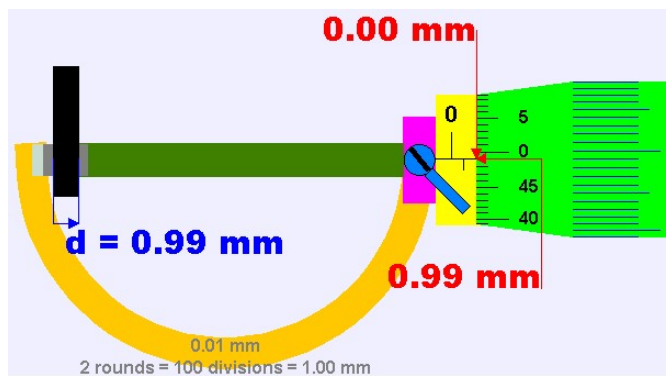

Figure 4. An EJS micrometer computer model with the micrometer dial (green) scale that has rotated slightly less than 2 rounds. The hints shows 0.00 mm on the main sleeve scale (yellow), 0.99 mm on the dial scale, giving the final answer of 0.99 mm. This model is downloadable https://dl.dropbox.com/u/44365627/lookangEJSworkspace/export/ejs_Micrometer02.jar.

Using the vernier caliper (Figure 1) and micrometer (Figure 2) to make measurements especially with the presence of zero error is a challenging task due to the difficulty of identifying exactly what to focus on, both on the vernier scale and the micrometer scale, and the amount of length to compensate for.

Thus, the customized model of a computer-simulated vernier caliper (Figure 3) and another of the micrometer (Figure 4) can complement the use of actual school laboratory vernier calipers and micrometers. This makes for greater authenticity in the learning experience, while enabling both educators and students to focus on the correct lines when taking readings. These two computer models share similar pedagogical design feature enhancements which will be discussed in the article. These computer models, also known as simulations, are created using a free authoring toolkit called Easy Java Simulation (EJS) [1] that we feel have excellent potential for advancing active experiential [2] student–centered learning for interactive engagement [3] through teacher customization [4] of computer models for their respective learning and teaching needs in schools.

In addition, our preliminary literature review using Google Scholar suggests that there is no prior documentation in *Physics Education* field, of the use of vernier caliper and micrometer simulation tools to help students learn the required GCE Ordinary Level Physics content and skills. Only two instances of documentation of other conference proceedings [5, 6] on this topic presented by the same author(s) of this article, are found.

This lack of scholarly research on vernier caliper and micrometer simulation could be due to the difficulties experienced in creating simulations with design features that are effective for learning and teaching. Building on Fu-Kwun work on source codes shared by their vernier caliper examples [7, 8] as well as micrometer examples [9, 10] these two Easy Java Simulation (EJS) [11, 12] customized computer models have been developed to serve as complementary virtual tools. These tools are downloadable from digital libraries ComPadre Open Source Physics (OSP) [13, 14] (simplified versions) and NTNUJAVA Virtual Physics Laboratory (login required). Alternatively, they can be downloaded using the following links https://dl.dropbox.com/u/44365627/lookangEJSworkspace/export/ejs_AAPTVernierCaliper.jar and https://dl.dropbox.com/u/44365627/lookangEJSworkspace/export/ejs_Micrometer02.jar. The recommended system requirement for running these EJS models is the Intel Pentium processor.

Step by step help in the use of these computer models is available on these YouTube videos http://youtu.be/zz4Fha5sISo and http://youtu.be/YAmn-xksu2s respectively.

## II. THREE COMPUTER MODEL PEDAGOGICAL DESIGN IDEAS

The following three computer model design insights that we found useful in our lessons hopes to add to the knowledge supporting the simulations' design features that afforded deeper learning through student directed inquiry. Readers interested in these design ideas can refer to other articles [2, 15, 16] in *Physics Education*.

### A. Simple 2D view for associated learning to the real world with consistent control panel

Contrary to the popular belief that 3D simulations could be more effective tools for learning, we discovered that for the context of vernier caliper and micrometer, a simple 2D view is more appropriate. Both models are designed to share similar consistent control panels (see Figure 5 and Figure 6) at the bottom of each simulations that are intended to make it easier for novice students to interact with the simulations. A zoom–in function is also added on the right panel that can be used to increase the visibility of the respective scales. This is especially useful during the educator's direct instruction using the projector screen of the models.

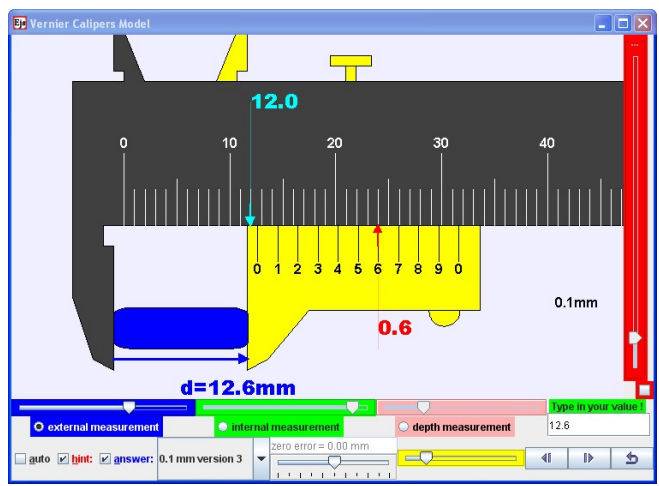

Figure 5. An EJS vernier caliper computer model that has a simple 2D view of the vernier caliper. The reading is with hint showing 12.0 mm on the top main scale, 0.6 mm on the bottom scale, giving the final answer of 12.6 mm. The bottom control panel allows different measurements (external, internal and depth), input field to check answers, different vernier scale versions and zero errors.

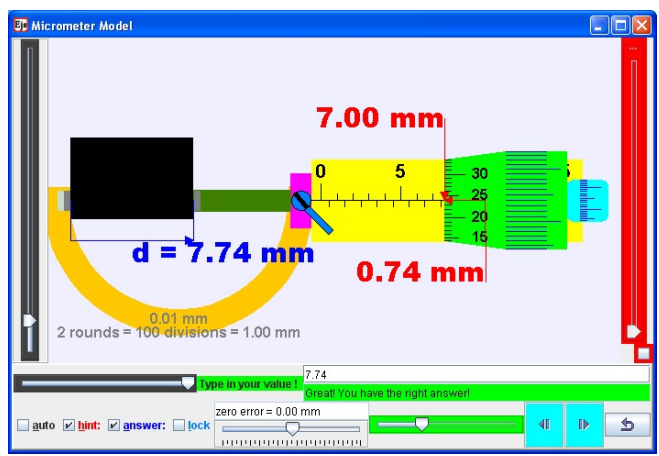

Figure 6. An EJS micrometer computer model that has a simple 2D view the micrometer. The reading is with hint showing 7.00 mm on the main scale, 0.74 mm on the (green) dial scale, giving the final answer of 7.74 mm. The bottom control panel allows a random sized external measurement, input field to check answers, (green) slider, (teal) thimble to take measurement and zero errors.

### B. Hints, answers, different scales and zero error

The hints and answers check–boxes (see Figure 5 and Figure 6) provide the 2 different scaffolding to understand the principles of the two measurement tools. In addition, the students are engaged in productive play as well as differentiated instructions through their own self-directed exploration to deepen their understanding.

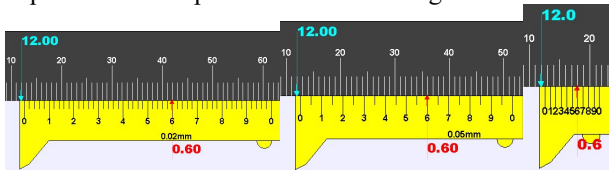

Figure 7. Different versions of the vernier scales from left to right are (0.02 mm, 0.05 mm and 0.1 mm) allows students to make sense of the vernier scale reading of 12.6 mm.

The different scale such as 0.02 mm, 0.05 mm and 0.1 mm options (Figure 7) can be customized to represent real vernier caliper tools that may be used.. In addition, the different (from left to right, 49 mm, 39 mm and 9 mm respectively) spanning vernier bottom scale length, can provide students with a wide variety of activities and hence is a valuable design idea for productive play [17].

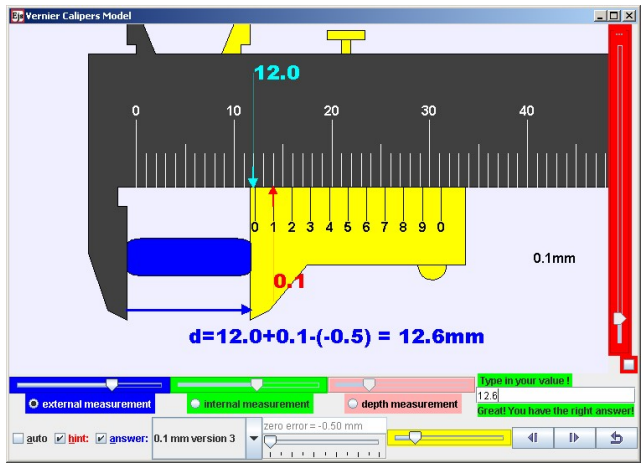

Figure 8. An EJS vernier caliper with the reading is with hints showing 12.0 mm on the top main scale, 0.1 mm on the bottom scale, with a zero error of – 0.5 mm, therefore, giving the final answer of 12.6 mm.

Lastly, the design idea of a variable positive or negative zero error option (Figure 8), has helped the students experience faulty measurement tools, a teachable moment that is rather difficult to achieve without damaging these real tools.

### C. Assessment for learning

We have designed these EJS models with input field and the customized (Figure 9) automated feedback to provide students with some fun, figuring out the logic of these computer models. Immediately after a submission through the input field followed by the "Enter" key on the keyboard, differentiated feedback messages can stimulate students to think and respond accordingly on subsequent action. Thus we believe this design idea has allowed students to practise, test and evaluate their understanding. Here, we recommend educators to discuss with their students on how to make sense of the real and virtual tools to help each and every student progress.

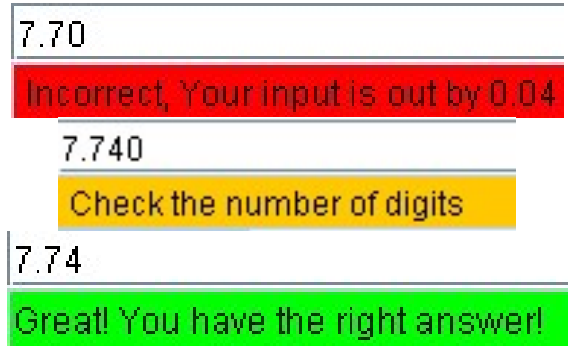

Figure 9. Input field with differentiated feedback such as "Incorrect, Your input is out by 0.04" a feedback to advise on the degree of inaccuracy by, "Check the number of digits" a feedback on the decimal places of precision, and "Great! You have the right answer!" to affirm the students' attempt to test his or her concept.

### III. PERFORMANCE OF LEARNING FROM STUDENTS

**Table 1:** Percentage of students (total number of student, 70) who selected choices (a) – (e) on Problem Questions Q(1) to (7) on the Google Site. The correct response for each question has been italicized. The average correct score was 88%

|   | Vernier caliper | | | | Micrometer | | |
| --- | --- | --- | --- | --- | --- | --- | --- |
| Q | 1 | 2 | 3 | 4 | 5 | 6 | 7 |
| (a) | 1 | 3 | *84* | 11 | 0 | 10 | 0 |
| (b) | 3 | *97* | 14 | 6 | 5 | 0 | 0 |
| (c) | *96* | 0 | 2 | *81* | *95* | 0 | 11 |
| (d) | 0 | 0 | 0 | 1 | 0 | 0 | 11 |
| (e) | 0 | 0 | 0 | 0 | 0 | *90* | *77* |

The 5E instructional model [18] guides the lesson design on the Google Site, where useful learning and teaching strategies have been incorporated.

Data from the student responses collected show that the students in this study scored above 96% correctly for Question 1 and 2 (Q 1,2) and 90% (Q 5,6) for simple reading of the rules (table 1) of vernier caliper and micrometer readings respectively.

With the introduction of zero errors, above 81% (Q 3,4) and 77% (Q 7) managed to get the correct answers to our 5 options (a) to (e) multi choice questions.

It is postulated that using direct instruction teaching after the student experiences both the real and virtual computer model can increase these percentages, emphasizing the rules as in equation (1).

Measure = main scale + vernier/dial scale – zero error   (1)

IV. FEEDBACK FROM STUDENTS

Excerpts from the qualitative survey results and informal interviews with the students reflect a positive learning experience. The data offers insights into the conditions and processes during the computer laboratory lessons. Words in brackets [ ] are added to improve the readability of the qualitative interviews.

*1) Interactive engagment [3] is fun and promotes learning*

"It was fun playing around with the computer model[s] and trying to read, then after that we could try measuring random things with the vernier caliper and micrometer screw gauge. Also, before this, I kept forgetting how to read the scale of these 2 tools, but this lesson has helped me remember better  :D [big smiley]".

"I think this lesson was good because it was a very different and enjoyable method of learning. It was also definitely helpful. I also found it more engaging compared to normal lessons, helping me stay focused. The pace of the lesson was also just right."

*2) Well designed lesson activities with real and virtual tools*

"This lesson helped me to refresh my memory of using the [real] vernier caliper and micrometer screw gauge. [T]he [learning] exercises given on the vernier [caliper] and micrometer hands on [with these real and virtual] gadget[s] were really useful in making sure we really understood how to use both [real vernier calliper and micrometer] gadgets."

"[This lesson is sucessful] because we got to see it in the real world and the virtual world. For example if we wanted to verify if it really happens in the real world, we can try it with the [real] vernier calliper we were given. Also, for the virtual vernier caliper, we could experiment with the different settings and see what would happen if we change one variable."

"It was an interesting way to learn about vernier caliper and micro[meter] screw gauge. It had both hands on and computer work to do which made it easier to remember how to use both. Also, it is something I have not done in a Physics lesson before."

*3)  well designed simulation [19]  with assessment for learning*

"It addresses all the uses of a vernier caliper. It also provided hints and answers to let us learn by ourselves without the teacher's help. When we have questions, we can refer to the hints and answers and try to identify the problem via our own eyes."

The survey results add to the existing understanding of a need to design assessment for learning through input field with customized feedback (Figure 9) for optimum cognitive loading. We postulate that the act of thinking and deciding on a measurement to key in the input field is not only making the students' thinking visible, but allow self-assessment of learning, not achievable with just observable hints and answers alone.

V. CONCLUSION

Two computer models are designed for teaching and learning purposes and we recommend their use as complementing actual vernier calipers and micrometers. These two computer models can be downloaded from http://weelookang.blogspot.sg/p/physics-applets-virtual-lab.html (no password nor login required), NTNU Virtual Physics Laboratory (login required) [11, 12]  and ComPadre Open Source Physics [13, 14] (simplified versions) digital libraries.

Three computer model design considerations such as 1) simple view with consistent control panel and associated learning to the real world, 2) hints, answers, different scales and zero errors and 3) assessment for learning for conceptual reasoning with, are implemented in our simulations that we believe can further support learning.

Feedback from the target group of students has been positive, triangulated from the performance of learning from 7 formative questions, survey questions, and interviews and reflections with students and discussions with teacher.

It is hoped that the 'success' of the pilot project  will encourage more teachers to use these simulations and publish their experiences in the science educators' community so that anyone may improve on these simulations for open education resources [20] advancing universal benefit for all.

ACKNOWLEDGEMENT

We wish to acknowledge the passionate contributions of Fu–Kwun Hwang, Francisco Esquembre and Wolfgang Christian, Doug Brown, Mario Belloni, Anne Cox, Andrew Duffy, Mike Gallis, Todd Timberlake, Taha Mzoughi and many more in the Open Source Physics community for their ideas and insights in the co–creation of interactive simulation and curriculum materials.

AUTHOR

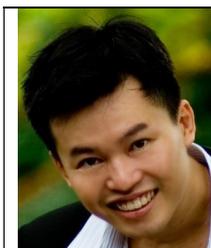
Loo Kang WEE is currently an educational technology specialist at the Ministry of Education, Singapore. He was a junior college physics lecturer and his research interest is in Open Source Physics tools like Easy Java Simulation for designing computer models and use of Tracker.

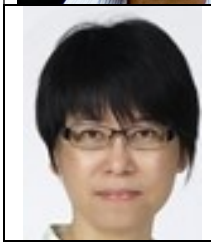
Hwee Tiang NING is currently a lead teacher at National Junior College, Singapore.

## VI. APPENDIX:

Q1: What is the reading of the vernier caliper shown below in mm?

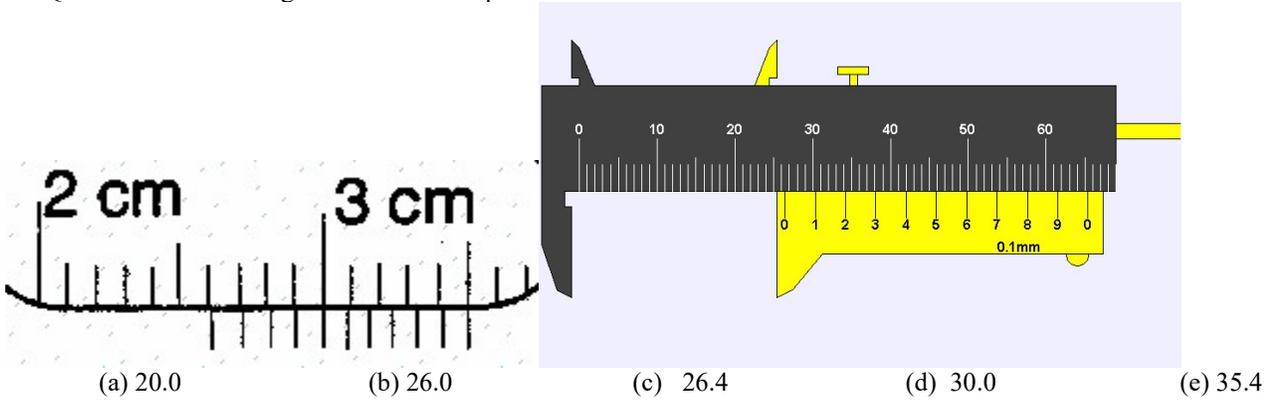

(a) 20.0      (b) 26.0      (c) 26.4      (d) 30.0      (e) 35.4

Q2 : The diagram shows a dice being measured using vernier caliper. What is the width of the dice, as recorded by the vernier scale?

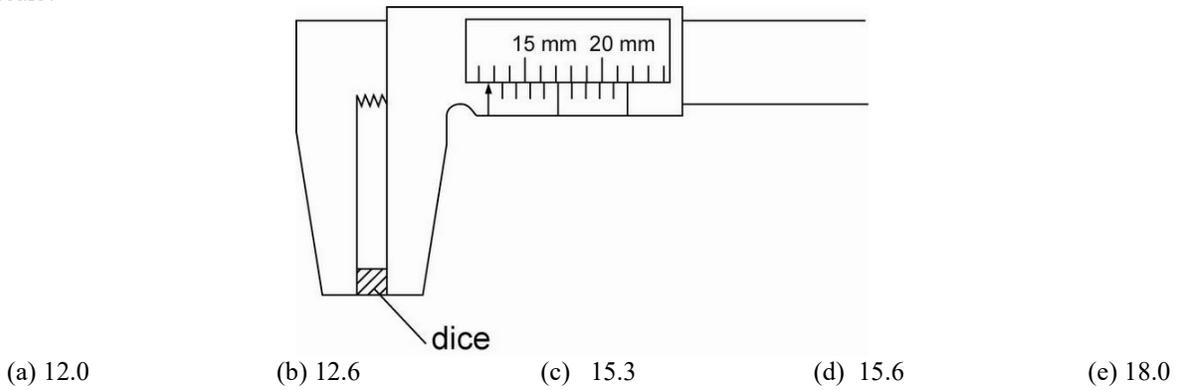

(a) 12.0      (b) 12.6      (c) 15.3      (d) 15.6      (e) 18.0

Q3: This vernier caliper has a Zero error = + 0.2 mm What is the correct reading of the vernier caliper shown below in mm (compuer model) ? and in cm ( pen paper diagram)

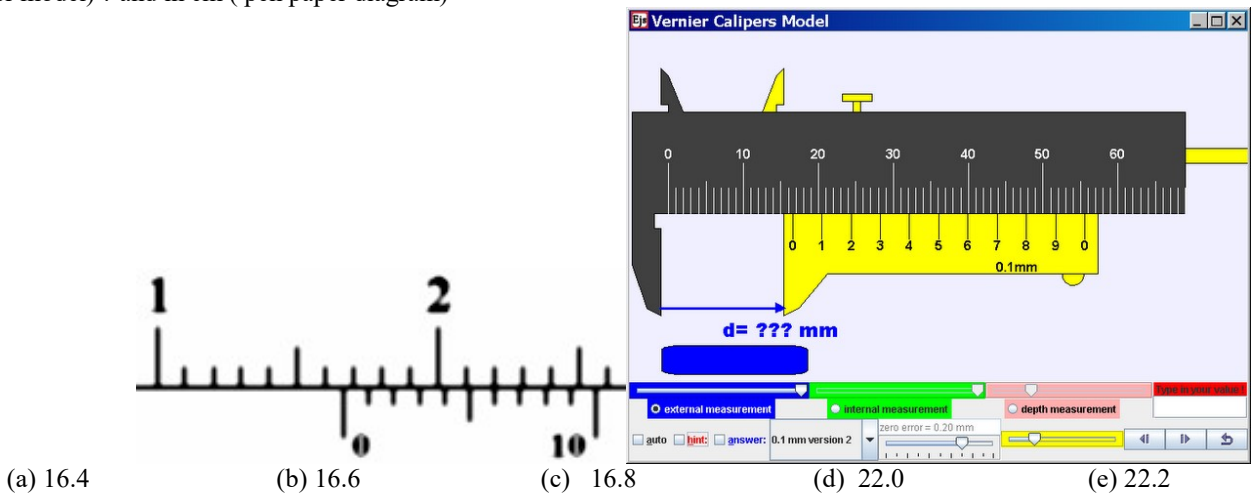

(a) 16.4      (b) 16.6      (c) 16.8      (d) 22.0      (e) 22.2

Q4: This vernier caliper has a Zero error = – 0.2 mm What is the correct reading of the vernier caliper shown below in mm?

(a) 16.4  (b) 16.6  (c) 16.8  (d) 22.0  (e) 22.2

Q5: What is the reading of the micrometer shown in mm?

(a) 5.25  (b) 7.24  (c) 7.74  (d) 8.24  (e) 8.74

Q6: What is the micrometer zero error reading shown in the diagram?

(a) –0.03  (b) 0.00  (c) 0.01  (d) 0.02  (e) 0.03

Q7: what is the correct reading on the micrometer given it has a zero error of – 0.03 mm?

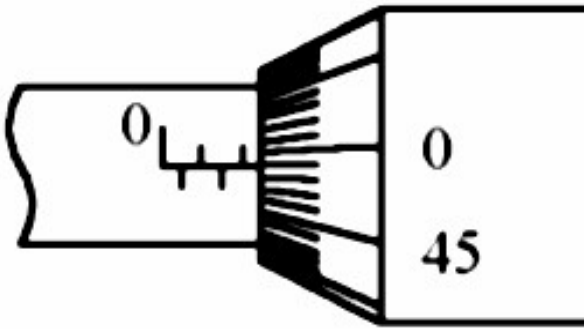
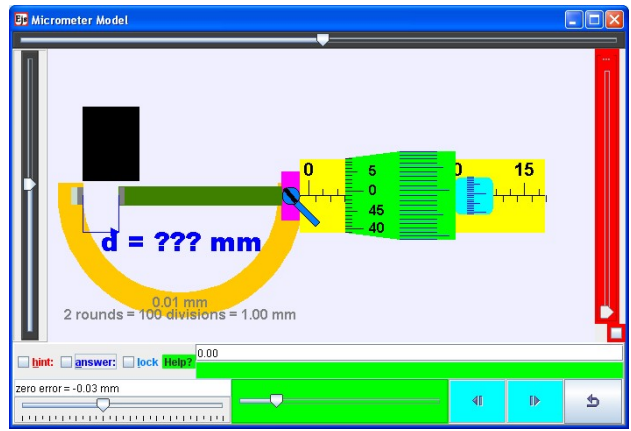

(a) 2.02  (b) 2.41  (c) 2.46  (d) 2.49  (e) 2.52